# The Early Years of Quantum Monte Carlo (2): Finite-Temperature Simulations

*Michel Mareschal[1,2],*

*Physics Department, ULB , Bruxelles, Belgium*

## Introduction

This is the second part of the historical survey of the development of the use of random numbers to solve quantum mechanical problems in computational physics and chemistry. In the first part, noted as QMC1, we reported on the development of various methods which allowed to project trial wave functions onto the ground state of a many-body system. The generic name for all those techniques is Projection Monte Carlo: they are also known by more explicit names like Variational Monte Carlo (VMC), Green Function Monte Carlo (GFMC) or Diffusion Monte Carlo (DMC). Their different names obviously refer to the peculiar technique used but they share the common feature of solving a quantum many-body problem at zero temperature, i.e. computing the ground state wave function.

In this second part, we will report on the technique known as Path Integral Monte Carlo (PIMC) and which extends Quantum Monte Carlo to problems with finite (i.e. non-zero) temperature. The technique is based on Feynman's theory of liquid Helium and in particular his qualitative explanation of superfluidity [Feynman,1953]. To elaborate his theory, Feynman used a formalism he had himself developed a few years earlier: the space-time approach to quantum mechanics [Feynman,1948] was based on his thesis work made at Princeton under John Wheeler before joining the Manhattan's project.

Feynman's Path Integral formalism can be cast into a form well adapted for computer simulation. In particular it transforms propagators in imaginary time into Boltzmann weight factors, allowing the use of Monte Carlo sampling. We will describe in the following how the Helium problem was formulated and put on the computer by David Ceperley and Roy Pollock in the eighties. By doing that, they transformed Feynman's theory of superfluidity into a quantitative theory being successfully tested against experimental data. They had to solve deep algorithmic issues which will be described below in a simplified manner.

---

[1] Retired

[2] email: mmaresch@ulb.ac.be

Ceperley and Pollock's work was not the first to try a path-integral approach to describe quantum systems in thermal equilibrium. However, it was the first computation which addressed and solved particle exchange sampling in addition to the necessarily biased sampling inspired from polymer physics. We will mention in this paper a few other works using PIMC and contemporary to the Ceperley-Pollock approach: none of them, however, needed to address the problem of particle exchange. Therefore, our main focus will be on the modelling leading to the Helium calculation.

The article is written at an elementary level, trying to be understandable by non-specialists. It is self-contained, in particular reading of QMC1 is not necessary. It also benefitted from the interview of David Ceperley that we have done in 2019 at the University of Illinois at Urbana-Champaign (UIUC). We will introduce the path-integral formulation at an elementary level: we believe that it is necessary in order to fully appreciate the difficulty of solving the problem on a computer. A detailed complete survey on the Helium computation has been published in a review article [Ceperley,1995].

In the following section, we briefly expose the challenge for building a theory of the lambda transition between normal and superfluid Helium phases. Next, we introduce the path integral formalism and make explicit the isomorphism between the quantum model and its classical counterpart. The computational challenges are then discussed in the two following parts, first the polymer sampling issue and second the problem of how to efficiently sample particle exchange. We end the article by placing the Helium computation into the context of other simpler PIMC approaches developed at around the same time. in the chemical physics community and we provide some short conclusions.

## The Helium Challenge

Helium, more precisely the He$^4$ isotope, has a unique phase diagram: lowering the temperature at low pressures, it remains liquid until zero Kelvin. It undergoes a phase transition at the so-called lambda temperature: below $T = T_\lambda = 2.2K$ a new form of liquid is observed. It has been called superfluid because it behaves as if there was no viscosity and no surface tension [Kapitza, 1938]. This transition is not observed with the other stable isotope of Helium, He$^3$, at least not at that temperature. Helium 4 is a boson, with total spin equal to zero, while Helium 3, with a neutron less in the nucleus, has a total spin 1/2 and therefore behaves as a fermion. Very quickly after the first observations, it was suggested by Fritz London [London, 1938] that the peculiar behaviour observed was linked to the quantum nature of the liquid. Bosons obey Bose-Einstein statistics and the theory predicts that lowering the temperature for an ensemble of non-interacting bosons leads to a phase change with an accumulation of all particles in the ground state: the so-called Bose-Einstein condensation. London imagined a theory where, below the transition temperature there would be coexistence between two fluids, one of them being made of the interacting atoms in the ground state.

Bose-Einstein condensation is easy to describe for a gas of free bosons. It was a real challenge for theoreticians in the late forties to extend the explanation to the case of liquid densities, to the point of matching experimental measurements of the transition. A theoretical model can be easily constructed as being made of an ensemble of particles, with Bose-Einstein statistics, and undergoing permanent and strong mutual interactions. Each helium atom can be considered as an elementary particle: electronic excitations require more than several hundred thousand Kelvin so that one can safely ignore internal structure and consider helium atoms as point massive particles, with a zero spin for Helium 4. Gas-phase experiments suggest that interactions between atoms are well modeled by short distance repulsion and long distance van der Waals attractive forces: very much like a two-body Lennard-Jones type of interaction[3]. The persistence of the liquid state up to zero Kelvin is due to a small attractive well, together with a relatively high zero-temperature kinetic energy.

Feynman proposed a microscopic theory of the superfluid transition in 1953 where he argued that "*the large interatomic forces will not alter the central features of Bose condensation* » [Feynman, 1953]. He used a formalism that he had previously developed [Feynman,1948], a review paper directly inspired from his PhD thesis which was dedicated to the Lagrangian formulation of quantum mechanics. With this formalism, he could map quantum mechanical particles in thermal equilibrium to a peculiar type of classical "polymers", providing a way to compute all static (i.e. non time-dependent) properties of the system. Feynman theory was in Feynman's style, intuitive and qualitative, so that many physicists remained doubtful about it: "*The physical idea which plays a central role is that, in a quantum-mechanical Bose liquid, atoms behave in some respects like free particles* »[Feynman, 1953]. Besides, he predicted the lambda transition being of third order whereas experiments were suggesting a second order (discontinuity of the heat capacity). Feynman's theory was later strengthened by a more mathematical approach [Penrose, 1956] but this latter work remained also qualitative. The lack of a precise calculation of the transition to superfluidity remained as a standing problem in low-temperature physics.

The advent of computer simulations and its success in dealing with classical liquids offered the possibility to revisit the path integral approach: Bose liquids can be mapped to classical systems which, in turn, can be simulated with either Monte Carlo (PIMC) or Molecular Dynamics (PIMD) methods. In principle, all static properties of a Bose superfluid can be calculated with no approximations on a computer.

## PIMC(1): From the quantum

In this section we will present the steps needed to go from the quantum expression of the free energy to the classical isomorphic model. While similar derivations are presented in various articles and textbooks (in particular see [Feynman,1955]), we present it here in a very simplified form to allow a more complete, self-contained, presentation. This part is however slightly more technical than the rest of the article and it can be omitted by readers who are less concerned in those aspects.

All physical properties of a set of $N$ quantum particles in thermal equilibrium can be obtained from the density operator: introduced by Von Neumann, the density operator generalises in the quantum case the phase-space probability density and allows the use of statistical ensembles. Knowledge of the equilibrium density operator, or density matrix leads to the so-called partition function, $Z$, which in turn provide the free energy, $F_N(V,T)$, and therefore all time-independent thermodynamic properties like the phase diagram, the heat capacity or the coefficient of thermal expansion, etc…

---

[3] Actual computations were done with a model of interactions proposed by Aziz [Ceperley,1995]; Aziz potential improves on Lennard-Jones in the modelling of Helium but is qualitatively similar.

$$e^{-\beta F(V,T)} = Z_N(V,T) = Tr\rho = \int dR \langle R|\rho|R \rangle \qquad (1)$$

where the representation of the density operator will be explicated below. Here, and below, symbols have their usual meaning, T being the temperature, etc.

The starting point of the analysis is a rather trivial identity for the canonical equilibrium density operator for a system with Hamiltonian $H$:

$$\rho = e^{-\beta H} = \left(e^{-\tau H}\right)^M \text{ with } \tau = \frac{\beta}{M} \text{ and } \beta = \frac{1}{k_B T} \qquad (2)$$

the quantity $\tau$ is called the time-step and it will be used as a smallness parameter. By analogy with the usual propagator $\exp\left(-\frac{iHt}{\hbar}\right)$, we can imagine following an imaginary time[4]. between 0 and beta, and $\tau = \beta/M$ would then be called the time-step.

Quantum effects are known to appear at low temperatures. More precisely the usual transition between classical and quantum behaviour occurs when the average distance between particles becomes of the order of the so-called de Broglie wavelength

$$\Lambda = \frac{h}{\sqrt{2\pi m k_B T}} \qquad (3)$$

Or, for a system of number density $n$, quantum effects will appear while lowering the temperature so that

$$n\Lambda^3 \simeq o(1) \qquad (4)$$

In the same way, beta will become large at low temperature but, for $M$ large enough, the timestep will still be considered as small and used as a smallness parameter. In particular, for a system with kinetic, $K$, and potential energy, $V$,

$$\begin{aligned} H &= K + V \\ e^{-\tau H} &= e^{-\tau K} e^{-\tau V} + o(\tau^2) \end{aligned} \qquad (5)$$

or

$$e^{-\beta H} = \lim_{M \to \infty} \left(e^{-\tau K} e^{-\tau V}\right)^M \qquad (6)$$

The writing of equation (5) is named the primitive approximation and is essential in the present analysis. It allows an explicit writing of equation (6) in a position representation. It is only exact when $M$ goes to infinity, which is known as the Trotter formula, but one expects the expression given in eq. (6) to be useful for $M$ sufficiently large.

---

[4] In Feynman's 1953 article: « *The quantity u is of course not the time. However, we shall obtain a vivid representation of (5) by imagining that it is the time"*. In Feynman's article, the quantity u=it/h goes from 0 to $\beta$.

One then can define a configuration variable, a *3N*-dimensional vector whose components are the positions of the *N* particles

$$R \Leftrightarrow (\vec{r}_1, \vec{r}_2 ... \vec{r}_N) \qquad (7)$$

In that position representation, one can evaluate the partition function, eq. (1). The potential energy is diagonal while the eigenfunctions of the kinetic energy are plane waves. So that, the density matrix elements can be written explicitly as the following expression:

$$\langle R_0 | \rho | R_M \rangle = \int ... \int dR_1 ... dR_{M-1} \frac{1}{(4\pi\Lambda\tau)^{3NM/2}} \exp\left\{-\sum_{i=1}^{M}\left[\frac{(R_{i-1}-R_i)^2}{4\Lambda\tau}\right] + \frac{\tau}{2}(V(R_{i-1}) + V(R_i))\right\} \qquad (8)$$

Equation (8) is the basic equation on which all PIMC will be based.

The path consists in the succession of configurations $R_0$, $R_1$..till $R_M$ which is followed in an imaginary time, step by step. For a diagonal element of the density matrix, the path is closed onto itself. The kinetic energy part of the Hamiltonian leads to gaussian factors between successive configurations, very much as if the path was made of successive transitions of a random walk. At each step one evaluates the Boltzmann weight of the interaction energy.

In equation (8), the argument of the exponential is called the action, *S*:

$$S(R_t) = \sum_{i=1}^{M}\left[\frac{(R_{i-1}-R_i)^2}{4\Lambda\tau}\right] + \frac{\tau}{2}(V(R_{i-1}) + V(R_i)) \qquad (9)$$

In classical mechanics, there is a minimum principle for the action. However, in order to compute the average in equation (8), all possible paths considered have wildly fluctuating values for the actions and they all contribute to the average.

The random walk interpretation of eq. (9) suggests a direct way to compute averages in a Monte Carlo simulation. One could start from a path, i.e. a succession of configurations

$P_{old} <=> \{R_0 => R_1 ... => R_{M-1} => R_M => R_0\}$

and imagine a random change of this path to a new path, for example by just changing the position of one particle at a given time step. We call this new path $P_{new}$. The new path would be accepted or refused depending on the following criterion, or probability of acceptance

$$A(P_{old} \Rightarrow P_{new}) = \min\left[1, \frac{e^{-S(R_{new})}}{e^{-S(R_{old})}}\right] \qquad (10)$$

By doing this a huge number of times, one would generate a large number of paths and therefore configurations on which one could compute average values of observables, at least those which can be expressed as functions of particle positions. Needless to say, the convergence of this naive approach is very poor in the general case.

# PIMC(2): to the classical model

Equation (8) can be viewed slightly differently: it can be interpreted as the classical partition function of a set of ring polymers, interacting in a peculiar way. To each particle would correspond a polymer made of *M* beads. Interaction between the beads inside a given polymer, is that of harmonic strings between nearest neighbours: this comes from the kinetic energy term of the Hamiltonian leading to the sum of position difference to the square in the exponential. In addition, interactions between beads belonging to different-polymer would take place only within the same time slice, i.e. between beads having the same $i$ index, ranging from *1* to *M*. An illustration is provided in figure 1, with 3 particles being represented.

$$\hat{\rho} = \left( e^{-\tau \hat{H}} \right)^M; \beta = \frac{1}{k_B T}; \tau = \beta/M$$

$$\langle \hat{O} \rangle = \frac{\text{Tr}[\hat{O}\hat{\rho}]}{\text{Tr}[\hat{\rho}]}$$

So, the mathematics of the previous section leads to an isomorphism between a quantum system made of *N* particles interacting through a potential function *V*, say Lennard-Jones type of interactions, and a classical system of *N* ring-polymer molecules, with harmonic strings between the *M* beads within the same chain, and the model interaction *V* between beads of different polymers having the same label inside the chain. One can figure out that the extension in space of this polymer is related to the de Broglie thermal wavelength and that quantum effects will appear whenever those ring polymers will have larger extension- this occurs at low temperature- and become intricated one inside the others.

$$Z_{R_1 \ldots} = \int dR_{1 \ldots} \langle R | e^{-\tau \hat{H}} | R_2 \rangle \langle R_2 | e^{-\tau \hat{H}} | R_3 \rangle \cdots \langle R_M | e^{-\tau \hat{H}} | R \rangle$$

$$= \lim_{M \to \infty} \left[ e^{-\tau T} e^{-\tau V} \right]^M$$

$$\ldots V(R_i) + V(R_{i+1}) \Big]$$

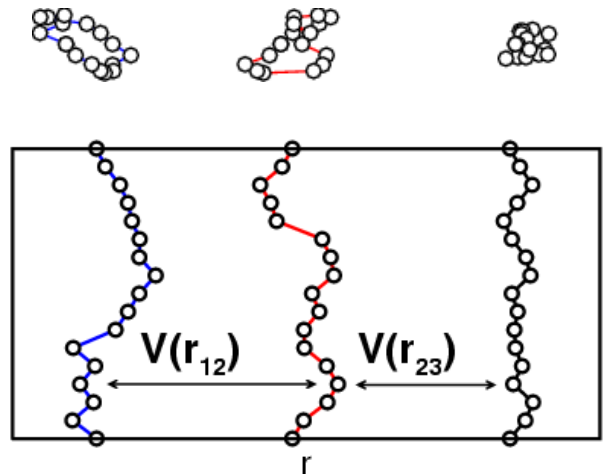

Figure 1: Classical isomorphim of 3 interacting quantum particles with M=17. In the upper part, the imaginary time is followed along a cyclic (ring) trajectory in real space, joining together the polymer beads. In the lower part, imaginary time is along the vertical axis; horizontal axis represents the space, possibly more than one-dimensional. Interactions between the beads take place within the same time slice, i. e. between beads at the same height.

To simulate the transition to superfluidity, however, one thing is missing which is related to the symmetry properties of the Helium atoms. We already argued that the transition occurs for the isotope of Helium having a zero spin and behaving as an ensemble of bosons. While eq. (8) would be sufficient for a system of « *Boltzmannons*" (i.e. distinguishable particles) , one needs an extra prescription to take into account symmetry (bosons) or antisymmetry (fermions) of the wave function. In other words, eq. (8) for bosons has to be made symmetric:

$$Z_N^{Bose}(V,T) = \frac{1}{N!}\sum_P \int dR_0....dR_{M-1} \exp\left[-\sum_{k=1}^{M} S^k\right] \tag{11}$$

In equation (11), the sum extends over all possible permutations over path closures: $PR_k=R_0$. This is the projection over symmetric states of the original density matrix. In other words, and this was Feynman's prescription, a path can close on another particle's position. For 2 particles, for example, one would have 2 contributions: one with two distinct polymers of length *M*, and another one with one polymer of length *2M*. An example is given in figure 2 for a 2-particle exchange. For 3 particles, one would have contributions from 6 different possibilities: (a) 3 distinct polymers of length *M*, (b) one polymer of length *2M* (dimer) and one of length *M* (3 possibilities) and (c) one polymer of length *3M* (a trimer: there are 2 distinct ways of realising this, given the two different particle orders which can be generated). And so on while increasing the number of particles. At high temperature, when the system has classical behaviour, the single polymers, « monomers" made of *M* beads, will dominate in the contribution to the partition function. Lowering the temperature, with an increasingly quantum behaviour, exchange, and therefore the longer polymers, made of 2, 3, 4… times *M* beads, will become more important in the sampling. Until at zero temperature polymers of all lengths become equally likely.

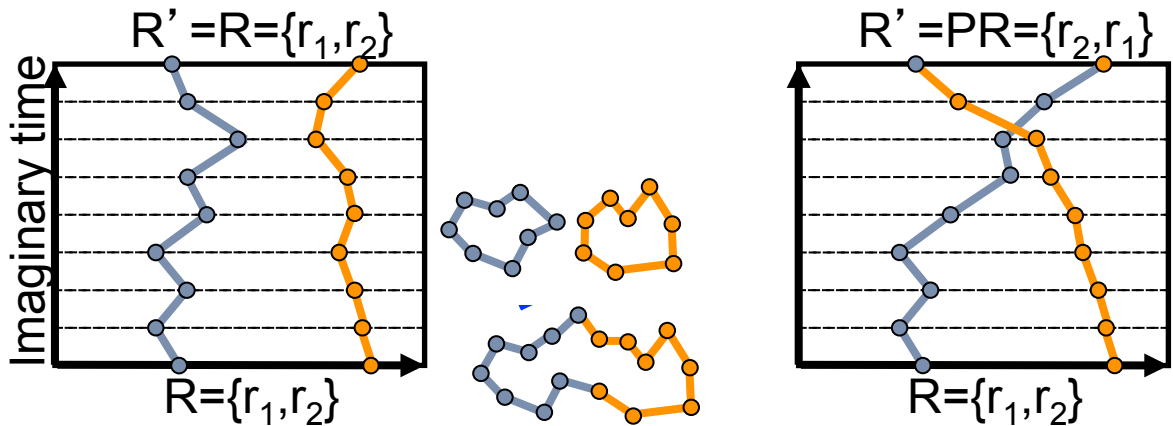

Figure 2: A 2-particle exchange by closing the path on another particle position: 2 *M=8* particles are exchanged and the result is a *M=16* ring polymer. The challenge is to propose a MC move going from left to right configuration (and inversely) with reasonable acceptance rate!

The isomorphism is now complete. The classical model to simulate would then consist of a set of polymers which would have a chemical equilibrium establishing the proportion of single, double, triple, …n-tuple rings as a function of the temperature. At high temperature, one would expect that single rings would be predominant, while lowering temperature, chains made of *nxM* beads would all become equally probable. The analogy with chemical equilibrium was emphazised in a remarkable article by Chandler and Wolynes [Chandler,1982]; the paper remained however somewhat formal as no practical way of implementing the technique was provided at that time. In the Monte Carlo computation, there is a need to introduce random moves with breakings and recombinations of polymers to allow for a chemical equilibrium to be established.

Another, earlier, work is worth being mentioned: it consists in a computational study using the path integral idea and involving quantum statistics to the point of proposing the sampling of particle exchange of non-interacting models. The study was achieved by Daan Frenkel, while visiting, as a graduate student, Gianni Iaccucci at Cecam during his PhD thesis. This study remained in the form of an unpublished letter joined to the 1975 report of activities of Cecam[5]. The paper convincingly shows the feasibility of using path-integral methods to compute thermal averages of quantum systems. It does probe the computation for simple models of non-interacting bosons and fermions , models for which a comparison with well-known exact results is possible and which do not need a large *M* value to show numerical convergence. The effect of the quantum statistics was shown in that report: the computed pair correlation function shows the effective attraction of bosons and the effective repulsion of fermions! The report argues on how to extend the method to interacting systems. In a private correspondance, Daan Frenkel explains the circonstances of this work and acknowledges the indirect role of John Barker[6]. It also refers to a previous work by Lawande and Salhin [Lawande,1969] on a similar topic, but for a small atomic model.

John Barker was a recognised expert in the theory of classical liquids: he was working on developing a method for finite temperature quantum mechanics, and eventually published a paper on quantum interacting hard spheres at finite temperature [Barker,1979]. His thinking has been influential for later developments: it has been acknowledged as motivating several works in the chemical physics community. In particular, when David Ceperley heard about the work just published by John Barker, he invited him for a seminar at Berkeley. Following the seminar, an intense discussion started with Roy Pollock, Malvin Kalos and Berni Alder who were also attending. This discussion convinced Ceperley and Pollock to dedicate further work on the quantum simulation of the Helium problem[7].

---

[5] Daan Frenkel, « *Feynman Paths and Quantum Monte Carlo* », in Rapport d'activité scientifique du CECAM, 1974-1975". Copy of the report can be found in the archives collected by CECAM at its headquarters at the EPFL in Lausanne.

[6] "*In the summer of 1975, Gianni Jacucci visited the US. He saw Berni Alder: I had asked Gianni to get me a hard-sphere code, but he was (of course) unsuccessful, so I wrote my own (this project overlapped with my path-integral work). Gianni also visited John Barker. When Gianni came back (in early September) he asked: "do you know about path integrals ? » I said "yes", because I had just read the book by Feynman and Hibbs. Then Gianni said: "John Barker is trying to compute quantum-virial coefficients, using path integrals. Wouldn't it be nice to use path integral Monte Carlo to simulate quantum many-body systems. I was enthusiastic and started right away. Actually, my plan was to do quantum dynamics by analytical continuation from the imaginary time axis. However, that was still too difficult in 1975. So, in the end we did simple Boltzmann, Fermi-Dirac and Bose Einstein ideal gases, and explored how to improve the sampling of interacting systems. However, at the end of September I had to return to Amsterdam, where I had little computing power (and I had to do experiments). So, the only thing that resulted is the attached report in the CECAM Annual Report 1975. The report acknowledges John Barker, and Bernie Goodman, whom I had never met. However, in 1981 Lee and Goodman published a paper on QMC of liquid Helium - but NOT using path integrals.*"

David Ceperley immediately understood the opportunity and the difficulty of the problem. He was then beginning employment at Lawrence Livermore National Laboratory where, with Berni Alder urging, he was completing the electron gas calculation. This work has since become a classic: the so-called Ceperley-Alder description of the electron gas used in many density functional approaches of first-principle molecular dynamics calculations within electronic structure models (see QMC1 for more details). Ceperley was then starting the hydrogen calculation with Berni Alder, the second fundamental application of his DMC (Diffusion Monte Carlo) technique. He had previously been working on classical polymer physics problems while achieving as postdoc in the group of Joel Lebowitz in New York and had therefore a clear idea of the difficulties due to the slow polymer dynamics in thermal equilibrium [Ceperley,1981].

Given his previous achievements and experience, David Ceperley was really in a unique position to undertake the PIMC computation of Helium. This calculation was an enormous challenge at the time, and it took several years to be achieved [Alder,1982;Pollock,1984;Ceperley,1986;Pollock, 1987]. It necessitated an intensive algorithmic effort to solve basically 3 different issues: (a) the classical isomorphic model for Helium atoms consists of ring-polymers of a peculiar type. Specific biased sampling schemes were definitely needed to obtain sensible equilibrium averages. (b) efficiency in the particle exchange sampling to be invented and (c) getting convergence with low enough values of *M*, the number of beads, to be compatible with the computer ressources available at the time (Cray1 and Cyber 205). The price to pay for this classical isomorphism is to have transformed a *N*-body problem into a *NxM* body problem, in addition to slow relaxation and a strange chemistry taking place.

## Computational Challenge (1) : improve the action!

The first simplification proposed was to simplify the action and it followed a suggestion made earlier by Barker [Barker,1979]!

It is worth also to mention that the limit *M*=> infinity of equation (8) leads to the Feynman-Kac formula, which can be written in the following way

$$\langle R|\rho(\beta)|R'\rangle = \langle R|\rho_0(\beta)|R'\rangle \left\langle e^{-\int_0^\beta d\tau V(R(\tau))} \right\rangle_{RandomWalk} \tag{12}$$

---

[7] Interview of David Ceperley: "*The work started when we became aware of work by John Barker (IBM San Jose) using path integrals for hard spheres. I organized a meeting with Barker, Pollock, Alder and Kalos in Berkeley in 1978 or 1979. He told us about his methods, in particular about the Metropolis procedure and the pair density matrix. While I was at Courant, Kalos had been working on a more complicated and less efficient scheme based on Green's function Monte Carlo[25]. Roy Pollock and I decided to go the route Barker had followed, using the Metropolis algorithm instead of the algorithm that Kalos advocated.*"

Comparing with eq. (8), the equation has been re-organized so as to make apparent the part due to the interactions between the particles: the bracket symbol represents an average over all paths undergoing random walks, starting at configuration *R* and ending in configuration *R'*.

In the models chosen for Helium, the interaction consists of a sum over all pairs of a potential function of a Lennard-Jones type. A great simplification occurs if we make the following approximation, interchanging average and product operations:

$$\left\langle e^{-\int_0^\beta d\tau \sum_{i<j} V_{ij}(R(\tau))} \right\rangle_{RW} = \left\langle \prod_{i<j} e^{-\int_0^\beta d\tau V_{ij}(R(\tau))} \right\rangle_{RW} \approx \prod_{i<j} \left\langle e^{-\int_0^\beta d\tau V_{ij}(R(\tau))} \right\rangle_{RW} \quad (13)$$

In the last expression, appears the interacting part of the exact action for a pair of atoms, a quantity which can be calculated exactly since it requires solving a 2-body problem [Storer,1969;Klemm, 1973].

$$\left\langle e^{-\int_0^\beta d\tau V_{ij}(R(\tau))} \right\rangle_{RW} = e^{-\beta u_2(\bar{r}_i, \bar{r}_j; \bar{r}_i', \bar{r}_j')}$$

This amounts to use the two-body density matrix in order to extract from it an averaged two-body action and to use this inter-action in the simulation. The practical way to do this is to solve numerically the two-body density matrix using one of the available techniques, for example the matrix squaring technique: the pair density matrix reduces to a function of 4 variables due to symmetry (3 differences in positions and one time difference) which can be stored and called during the simulation.

The use of the exact two-body density matrix allows to obtain much faster convergence in the whole by reducing the number of time slices needed by two orders of magnitude. For Helium at the lambda transition state, a value of *M* around 50 was sufficient to obtain stable results.

There are many arguments sustaining the above approximation. Let us cite one of them: the approximation is exact a low density and very good at zero-temperature as it tends to the Jastrow pair product function which describes the ground state so well. The approximation amounts to neglect 3-body, 4-body correlations in the description of the interaction between two particles: the correlation of two particles become independent of the others during a short time interval."*It is similar to integrating the solvent degrees of freedom in the description of interacting solute molecules* ». Let us stress that the average made is over the quantum dispersion of the values taken by the action along the paths joining *R* and *R'*.

The other algorithmic solutions concern improving sampling efficiency, both the sampling of polymer chains and the sampling of exchange.

# Computational Challenge (2) : biased Monte Carlo

One can easily figure out that ring-polymer peculiarities, the harmonic springs becoming stiffer with *M*, the way by which the interactions take place and the polymerisation of ring-polymers- do not allow for a straightforward application of Monte Carlo techniques which have been developed for simple liquids.

The simulation of classical liquids on computers was made possible by the finding of the so-called Metropolis algorithm in a historic paper published in 1953 [Metropolis, 1953]. The algorithm consists in exploring the configuration space of a classical system by realising a random walk with rules such that averages would rapidly converge to canonical averages. The algorithm is simple to describe: one starts from a low-energy configuration of a system of classical particles modelling, say, a liquid: for example point particles, interacting through a Lennard-Jones potential, and arranged on lattice nodes. Then, using random numbers, one makes a random move, say one particle randomly chosen is displaced at random in a cube centered on its original position. The new position is accepted if the interaction energy, *U*, goes down. Otherwise, it is accepted with probability $\exp[(U(\text{old})-U(\text{new}))/k_B T]$.

$$P(old \rightarrow new) = \min\left\{1, \exp\left[\beta\left(U(new) - U(old)\right)\right]\right\} \qquad (14)$$

If the move is refused, the old configuration is counted as a new one and the process is repeated. The algorithm belongs to the class of stochastic processes known as Markov chains for which a theory had been developed [Feller, 1957]; it can be shown that it samples configurations according to the canonical equilibrium distribution. Besides, at least in the case of simple models, it converges rapidly: average values of observables computed on those generated states rapidly tend to their equilibrium averages.

The random move to go from the old configuration to the new one can be arbitrarily chosen with a criterion of convenience and efficiency, with the only restriction that it should obey detailed balance: this last requirement is crucial to generate equilibrium states. By detailed balance, it is meant that, if one can move the system from one configuration to another it should be possible to have the reverse move as well. At equilibrium one should have equal (equilibrated) flows of probabilities between the two configurations.

$$P(old \rightarrow new) w_{old}^{eq} = P(new \rightarrow old) w_{new}^{eq} \qquad (15)$$

In the case of the Metropolis algorithm, it is easy to verify that detailed balance is obeyed by using equation 14.

Already in 1955, Marshall Rosenbluth [Rosenbluth,1955], who had been at the origin of the Metropolis algorithm, proposed a slightly modified procedure in the case of polymer modelling: he wanted to grow a polymer on a lattice and therefore he wanted to favour most likely configurations, avoiding loosing computer time in calculations for unlikely moves. Instead of choosing a random move for the new bead to continue the polymer growth, he biased the choice in favour of selecting the most likely positions. Let us call *f[U(new)]* the probability to choose the new state, depending on the energy of the new state with the already grown polymer, then a possible acceptance probability of the new state could be:

$$P(old \rightarrow new) = \min\left\{1, \frac{f(U(old))}{f(U(new))}\exp\beta\left[U(new) - U(old)\right]\right\} \qquad (16)$$

In other words, one can bias the choice for the new state, enhancing acceptance of the moves. The choice can be based on efficiency, so that the probability of acceptance is sufficiently high to allow for a reasonable computing time. The acceptance probability in equation 6 obeys detailed balance and the whole process remains within the safe and known domain of Markov chains.

Rosenbluth sampling has since been generalised and has led to various Monte Carlo techniques dealing with complex liquids: growing or moving complex molecules in such a way that a sufficiently high fraction of the new positions in the random walk will be accepted and correcting in the computed averages. The technique has been named *biased Monte Carlo* and has been become part of standard Monte Carlo techniques for complex liquids [Frenkel, 2002].

The problem encountered by Ceperley and Pollock share some aspects with the Rosenbluth case in as much as the problem of ring polymers is a problem of polymer relaxation. Ceperley had been doing a postdoctoral research while staying at Rutgers in 1979, leading to several papers on the statistical mechanics of polymers [Ceperley, 1981]. First, Ceperley and Pollock quickly realised that the move to perform should involve several beads of a ring polymer, given the stiffness of the polymer at larger $M$. They could determine how the centre of mass of the chain would diffuse during a simulation where only single-slice moves would be performed: they found it would scale like $M^{-3}$. An increase by a factor ten of the number of beads would slow down simulation time by a factor one thousand.

The sampling efficiency improves much with collective moves of several beads of the whole chain: if there was one single chain, it would be possible to propose moves inspired by the one-particle density operator which can be easily calculated and gives basically gaussian probability distribution of relative bead distances which can be used to grow the polymer. This led to a technique known as the *bisection* method: they cut a piece of polymer of a few beads (say 8) and regrow it, by first choosing the mid-point between the two free-ends, then the other two mid-points and so on. Testing acceptance at each step of the process leads to less computations being made in case of unsuccessful moves. The basic idea consists of attempting moves with computation of the gross features before sampling finer details. Other similar algorithms to achieve this were made in the so-called *staging algorithm* [Sprik, 1985], or in methods inspired by DMC.

The possibility to have moves involving a substantial part of a ring-polymer is not only much more efficient to explore configuration space, it is also essential when considering exchange.

# Computational Challenge (3): permutation sampling

The other important idea developed by Ceperley and Pollock was to sample permutation space. Indeed, in the exact expression for a $N$ particle system, there are $N!$ terms: the number of permutation among the N particles. The magnitude of this number makes it impossible to calculate the sum. But all terms in the sum are positive (unlike for fermions where one has positive and negative values). Sampling permutation space means that the boson simulation consists in a random walk through the (discrete) permutation space, in addition to the random walk into configuration space.

We have already mentioned the possibility for ring polymers to cross link: in an equilibrated system, one should have contributions coming from polymers made of *M* beads, *2M* beads, *3M* beads and so on. In the sampling, one should therefore allow for a kind of chemical equilibrium between n-tuple polymers: breaking and reforming linking of *n times M* polymers. Note that the time slice where this occurs is arbitrary: permutation can occur at any time slice as it would be equivalent to a change of integration variables.

The way to achieve this consists in establishing a permutation table: list all pairs of particles which are sufficiently close so that one can realise a move consisting in cutting two pieces of the ring-polymers (say 8 beads) and rebuilding the links in order to change the existing cross-linking. The move is very artificial but it should allow for either building a crosslink between two separate particles, or, in a reversible way, de-construct an existing link. The reversibility of the change is essential for detailed balance and the establishment of thermal and chemical equilibrium. The way to create and annihilate links between polymers is by using the bisection method described in the previous section (or any alternative to it).

To reach better efficiency, Ceperley and Pollock also included in the list possibility of three-particles permutations, since very often, while considering a pair of neighbours, a third particle can stand in-between. Since, the Monte Carlo move needs not be following any physical law, any breaking and building of links can be imagined, provided it allows for an inverse move, so that detailed balance is respected. In some cases, direct moves can also be considered, breaking or building links with no physical move of beads, but those permutations are rare compared to the ones which can be realised with a rebuilding of the beads positions. With the combination of both the particle exchange and the chain reconstruction, the code could generate a few cross-linking moves every second on a Cray-1 supercomputer, which allowed for computing the thermodynamics of the helium 4 model.

# Confronting experiments and …theory!

We have been very sketchy in the description of the algorithms used in order to sample the classical system of ring-polymers. A very detailed description was published in 1995 by David Ceperley in an article for the Review of Modern Physics [Ceperley, 1995]. This article is a deep and complete exposition of the work done on helium, going into the details of both the algorithmic efforts as well as the physical results obtained. The results were first obtained during the eighties, leading to comparisons with experimental data. The lambda transition itself could be reproduced, with the peak in the heat capacity: the results appeared in an article of Ceperley and Pollock [Ceperley, 1986] and comparison was made with the best existing data at the time [Wilks, 1967]. Of course, because of the finiteness of the simulated system, a smoothing occurs in the transition but the comparison with experiment is impressive.

All the static properties of liquid helium could be computed and compared to experiments: pair correlation functions, structure factors were compared to experimental data coming from X-ray and neutron scattering. It was clear that the method was providing exact theoretical results for static properties of quantum liquids, as it was already the case for classical computational liquid theory.

Some results were more difficult to obtain. For example, the momentum distribution, which is a good indication that the Bose-Einstein condensation takes place. The condensate fraction was obtained by the calculation of the end-to-end distance of a polymer. More precisely, the condensate fraction turns out to be the ratio of time the polymer is stretched out to the time it is coiled up during the whole random walk. It is a collective property and this is linked to the uncertainty principle. Determining the momentum of an atom makes the polymer delocalised in space: it corresponds to an open polymer. In normal helium, at high temperature, the polymers are coiled up. Below the critical temperature (2.2 K), long permutations become frequent and the two ends of a polymer spontaneously become separated from each other. Using open paths allowed a direct computation of the momentum distribution, and therefore of the condensate fraction as a function of the temperature. This was new with respect to Feynman's papers.

Another theoretical result which came out of understanding the simulation was the relation of the superfluid density to a quantity known as the winding number. The winding number, *W*, characterises paths which are winding around the simulation cell: because of periodic boundary conditions, a path can disappear on one side of the cell and reappear from the opposite side: a few paths are winding around the cell. Taking this into account this boundary term was necessary to obtain a meaningful (and correct) result for the superfluid density, i.e. the fraction of the system that responds to an imposed motion. This led also to deriving an exact formula for the superfluid density, published in [Pollock, 1987]

$$\frac{\rho^s}{\rho} = \frac{2mk_B T \langle W^2 \rangle}{\hbar^2 N} \qquad (17)$$

This was shown to be an exact formula, providing the connection between long exchanges and the superfluid density. The formula was derived after a direct inspection of the simulations, it had been missed previously. As emphasised by David Ceperley: «*I think there are many similar formulas still to be discovered in Quantum Monte Carlo. They're not in the textbooks because people haven't worked on QMC to the extent of other subjects. This is one of the times that looking at the simulation led to a much deeper theoretical result and something that has important implications for simulations. With the calculation of the superfluid density, we had explained many of the unusual features of superfluid helium starting from the interatomic potential. No adjustable parameters or uncontrolled approximations were needed. I think the PIMC has had an even larger impact on theoretical physics than the zero temperature QMC calculations, since it is more closely related to classical statistical mechanics* »

David Ceperley mentions also that he had an occasion to present his results to Feynman in 1983. As he mentions it :" *It was during that period (Jan. 1983) that I met Richard Feynman and had a chance to describe to him a computer implementation of his path integral theory of superfluid helium. Werner Erhart, the head of EST (Erhart Seminar Training, a "human potential movement" based in San Francisco, and dissolved in 1984) wanted to have a workshop that would attract a group to talk about the frontiers of science, in this case simulation of quantum systems…I was just starting to get results on liquid helium using PIMC, so my talk was on DMC results for the electron gas and hydrogen. Feynman was interested in how to apply the renormalization ideas that Swendson had discussed to the superfluid transition. We discussed the relation of the permutation distribution to the superfluid transition. This was about the time that Feynman was interested in*

*quantum and parallel computing. During the workshop, he was clearly was interested in what was going on in quantum simulation and whether classical computers could tackle the quantum world.*"

## Other PI works

In this paper, we have focused the description of the progress achieved on Helium by Ceperley and Pollock among all works in finite-temperature QMC. The reason being that this work was confronted at the same time with the difficulties of sampling polymer configurations and particle exchange. Other works[8] were reported in the eighties in QMC. For example, David Chandler and Peter Wolynes published a paper based on Feynman'ideas on path integrals and could formulate their approach up to the point of using results from polymer's chemical equilibrium theories. This approach also introduced the concept of isomorphism: however it did not go to the point of proposing actual ways of performing the samplings and remained somewhat formal. Later on, David Chandler applied his ideas to the theory of electron transfer in solution where he could describe and compute the ferrous-ferric equilibrium : Chandler's work on ferrous-ferric electron transfer in water [Kuharski,1988] could confirm that the electron transfer is well described by a spin-boson model, in which a two-state system is linearly coupled to a harmonic bath. This work was a confirmation from first principles of the Marcus theory of electron transfer which led to a chemistry Nobel prize in 1992: Chandler's work was cited in the Nobel lecture by Rudy Markus.

The same absence of permutation sampling was true for another powerful approach in chemical physics: one electron in solution was described by Klein and Sprik, joined at some point by David Chandler. In that work they developed the staging algorithm to perform biased Monte Carlo moves for polymer sampling. At room temperature, the electron is delocalised over hundreds of classical solvent molecules which can be treated classically. This leads to high values of $M$, the number of beads of the so-called polymer, which can reach values of a few thousands. Efficient sampling does not occur spontaneously in those stiffed polymers and the so-called "staging" algorithm which has been proposed [Sprik,1985; 1988] is achieving a task equivalent to the so-called bisection method of Ceperley.

One should also mention the work by Anees Rahman and Michele Parrinello [Parrinello, 1985]. PIMC was used to perform a mixed quantum-classical simulation: a PIMC description of a quantum object, a charged fluor atom, is embedded into a classical environment, with no need to solve issues linked to particle exchange.

Besides Monte Carlo, there appeared during the eighties techniques using Molecular Dynamics [Tuckerman,2002] rather than Monte Carlo to perform the sampling in the configuration space of the isomorphic classical model. One could naively think that this approach would be superior to the Monte Carlo, as it would also provide the dynamics of the system. However, this is not so: the isomorphism is limited to the static properties, the dynamics of PIMD is artificial: despite clever algorithms developed to achieve efficient sampling no time-correlations can be computed within this MD approach.

---

[8] The proceedings of a workshop organised in december 1982 in Paris by Malvin Kalos are a valuable report on the state of the art in that period: see [Kalos, 1982]; see also [Sauer, 2015]

# Conclusions

In the introduction of his book on path integrals and quantum mechanics [Feynman,1965], Feynman argues that the formalism developed is of pedagogic use: « *At the present time, it appears that the operator technique is both deeper and more powerful for the solution of more general quantum-mechanical problems* ». While this appreciation is possibly true in the case of analytical work, it is certainly not so in computational approaches. As a matter of fact, what appears from inspecting the literature is that the path-integral formalism provides a unique way of handling quantum mechanical statistical averages at a finite temperature on a computer. The technical reason for this lies in the fact that the propagator with imaginary time can be cast in a form of a Boltzmann weight factor, which, in turn, can be efficiently evaluated with the Metropolis Monte Carlo techniques developed in liquid theories.

Nevertheless, the problem remains intrinsically extremely complicated. The Helium 4 PIMC computation was a real *tour de force:* achieved in the eighties: it was done on the biggest supercomputers of the time, taking advantage of the large computing time available in Alder's group at Livermore. It also required to create new algorithms to achieve efficient sampling of configurations and particle exchanges.

Since then, progresses have been constant in the hardware as well as in the algorithms, addressing more complex models. Those early computations are nowadays looking like routine in terms of computer requirements. The PIMC method has since been extended to more complex cases with fermions being considered as well as bosons and with no limitations in the phase diagram of simple (i.e. light) materials [Pierleoni,2006; McMahon,2012]. Looking retrospectively at those developments, it is clear that the He4 computation has been an important breakthrough in the computer modelling of matter, opening several new ways of further investigation.

# Acknowledgements

The idea to undertake this study was a suggestion by Giovanni Ciccotti whom I wish to thank for a constant support, a generous availability and many useful suggestions. I wish also to thank David Ceperley for his hospitality and patience during my visit to him in Illinois and for later correspondance. Stimulating discussions with Benoit Roux are gratefully acknowledged.
David Ceperley's interview was made possible through financial support provided by the Neubauer's Collegium of the University of Chicago.

During the completion of this article came the sad news of the death of Berni Julian Alder the day preceding his 95th anniversary. I wish to acknowledge here his constant, generous and enthusiastic availability for sharing his memories as well as his views on the field he has contributed to create.

# Bibliography

ALDER, Berni J., CEPERLEY, David M., and POLLOCK E.L., *Computer Simulation of Phase Transitions in Classical and Quantum Systems*, International Journal of Quantum Chemistry **16**, 49 (1982).

BARKER John A. , *A Quantum-statistical Monte Carlo Method: path integrals with boundary conditions* , J. Chem. Phys. **70**, 2914-2918 (1979)

CEPERLEY David M., KALOS Malvin.H., and LEBOWITZ Joel L., *The Computer Simulation of the Static and Dynamic Properties of a Polymer Chain*, Macromolecules **14**,1472 (1981).

CEPERLEY David M., and POLLOCK, E.L., *Path Integral Computation of the Low Temperature Properties of Liquid $^4$He*, Phys. Rev. Lett. **56**, 351 (1986).

CEPERLEY David M., and POLLOCK, E.L., *The Momentum Distribution of Normal and Superfluid Liquid 4He*, Can. J. Physics **65**, 1416 (1987).

CEPERLEY David M., *Path Integrals in the Theory of Condensed Helium*, Rev. Mod. Phys. **67**, 279 (1995).

CHANDLER David and WOLYNES Peter G., *Exploiting the isomorphism between quantum theory and classical statistical mechanics of polyatomic fluids*, J. Chem. Phys. **74**, 4078-4095 (1981)

FELLER, William, *An Introduction to Probability Theory and its Applications,* vol. 1, John Wiley and Sons, New York (1957)

FEYNMAN Richard P., *Space-Time approach to non-relativistic quantum mechanics*, , Rev. Mod. Phys. **20**, 367-387 (1948)  https://doi.org/10.1103/RevModPhys.20.367

FEYNMAN Richard P., *Atomic Theory of the lambda transition in Helium*, Phys. Rev. **91**, 1291-1301(1953)

FEYNMAN Richard P. and HIBBS Albert R., *Quantum Mechanics and Path Integrals*, MacGraw-Hill, New York (1965) and emended edition by David F. Slyer (2005)

FRENKEL Daan and SMIT Berend, , *Understanding Molecular Simulations: from Algorithms to Applications*, Academic Press, Elsevier, 2002 doi.org/10.1016/B978-0-12-267351-1.X5000-7

KALOS Malvin H. ,*Monte Carlo Methods in Quantum Problems*, M. H. Kalos editor, NATO proceedings 125, 1982.

KAPITZA, Piotr, *Viscosity of Liquid Helium below the lambda Point*, Nature **141**, 74 (1938)

KLEMM A. D. and STORER R.G., *The Structure of Quantum Fluids: Helium and Neon*, Austr. J. Phys. **26**, 43-60, 1973.

KUHARSKI RobertA., BADER Joel S., CHANDLER David, SPRIK Michiel and KLEIN Mike L., *Molecular Model for Aqueous Ferrous-Ferric Electron Transfer*, J. Chem. Phys. **89,** 3248–3257 1988.

LONDON Fritz, *On the Bose-Einstein Condensation*, Phys. Rev. **54**, 947-954 (1938)

LAWANDE Suresh V., JENSEN C. A. and SAHLIN H. L., *Monte Carlo Evaluation of Feynman Path Integrals in Imaginary Time and Spherical Polar Coordinates*, J. Comp. Phys. **4**, 451-464 (1969)

McMAHON, Jeffrey M., MORALES Miguel A., PIERLEONI Carlo and CEPERLEY David M., *The Properties of Hydrogen and Helium under Extreme Conditions*, Rev. Mod. Phys. **84**, 1608-1653 (2012). DOI: 10.1103/RevModPhys.84.1607

METROPOLIS, Nick, ROSENBLUTH Ariana W., ROSENBLUTH Marshal N. , TELLER A. and TELLER Edward, , *Equation of State Calculations by Fast Computing Machines*, J. Chem. Phys. **21**, 6, 1087-1092 , 1953

PARRINELLO Michele and RAHMAN Anees, *Study of an F Center in Molten KCl*, J. Chem. Phys. **80**, 860-867 (1984). doi.10.1063/1.446740

PENROSE Oliver and ONSAGER Lars, *Bose-Einstein Condensation and Liquid Helium*, Phys. Rev. **104**, 576 (1956). doi.org/10.1103/PhysRev.104.576

PIERLEONI, Carlo and CEPERLEY David M., *The Coupled Electron-ion Monte Carlo Method*, in *Computer Simulations in Condensed Matter Systems: from Materials to Chemical Biology*, eds. M. Ferrario, G. Ciccotti and K. Binder, Lecture Notes in Physics, vol. **703**, 641-683, Springer Berlin Heidelberg (2006)

POLLOCK, E.L., and CEPERLEY David M., *Path-Integral Computation of Superfluid Densities*, Phys. Rev. **B 36**, 8343 (1987).

POLLOCK, E.L., and CEPERLEY, David M., *Simulation of Quantum Many-Body Systems by Path Integral Methods*, Phys. Rev. **B 30**, 2555 (1984).

ROSENBLUTH Marshall N. and ROSENBLUTH Ariana W., *Monte Carlo Calculation of the Average Extension of Molecular Chains*, J. Chem. Phys. **23**, 356 (1955)

SAUER Tilman, *The Feynman Path goes Monte Carlo*, in *Fluctuating Paths and Fields. Festschrift Dedicated to Hagen Kleinert*, Eds. W. Janke, A. Pelster, H.-J. Schmidt and M. Bachmann, World Scientific , Singapore, 2001.

STORER R. G., *Path-integral Calculation of the Quantum-Statistical Density Matrix for Attractive Coulomb Forces*, J. Math. Phys., **9**, 964-970 1968 doi.org/10.1063/1.1664666

SPRIK Michiel, KLEIN Michael L. and CHANDLER David, *Staging: A sampling technique for the Monte Carlo evaluation of path integrals*, *Phys. Rev.* **B 31,** 4234–4244 1985

SPRIK Michiel and KLEIN Michael L., *Application of path integral simulations to the study of electron salvation in polar fluids,* Comp. Phys. Rep. **7-3**, 147-166 (1988). dos.org/10.1016/0167-7797(88)90001-9

TUCKERMAN Mark E., BERNE Bruce J., MARTYNA Glen J. and KLEIN Michael L., *Efficient Molecular Dynamics and Hybrid Monte Carlo for Path Integrals*, J. Chem. Phys. **99-4**, 2796-2808 (1993)